\documentclass[12pt,preprint]{aastex}
\usepackage{graphics,epsfig}
\newcommand{\etal}{et al.~}

\begin{document}

\title{Serendipity and the SDSS: Discovery of the Largest Known Planetary
Nebula on the Sky}

\author{Paul C. Hewett,\altaffilmark{1} \email{phewett@ast.cam.ac.uk}
Michael J. Irwin,\altaffilmark{1} \email{mike@ast.cam.ac.uk}
Evan D. Skillman,\altaffilmark{1,2} \email {skillman@astro.umn.edu}
Craig B. Foltz,\altaffilmark{3} \email{cfoltz@nsf.gov}
Jon P. Willis,\altaffilmark{4,5} \email {jwillis@uvic.ca}
Stephen J. Warren\altaffilmark{6} \email{s.j.warren@imperial.ac.uk}
and Nicholas A. Walton\altaffilmark{1} \email{naw@ast.cam.ac.uk}}

\altaffiltext{1}{Institute of Astronomy, Madingley Road, Cambridge, 
CB3 0HA, UK}

\altaffiltext{2}{Department of Astronomy, University of Minnesota, 116 Church Street, 
SE, Minneapolis, MN 55455}

\altaffiltext{3}{Division of Astronomical Sciences,
National Science Foundation, Room 1045, 4201 Wilson Blvd, Arlington, VA 22230}

\altaffiltext{4}{European Southern Observatory, Alonso de Cordoba 3107, 
Vitacura, Casilla 19001,
Santiago 19, Chile}

\altaffiltext{5}{Current address: Department of Physics \& Astronomy, University
of Victoria, PO Box 3055 STN CSC, Victoria, BC, V8W 3P6 Canada}
\altaffiltext{6}{Astrophysics Group, Blackett Laboratory, 
Imperial College London, Prince
Consort Road, London SW7 2BW, UK}

\bigskip

\begin{abstract} 

Investigation of spectra from the Sloan Digital Sky Survey reveals the presence
of a region of ionized gas of $>2\arcdeg$ diameter centered approximately at
$\alpha = 10^{h} 37^{m}$ $\delta = -00\arcdeg 18'$ (J2000) (Galactic coordinates $l=248, b=+48$).
[\ion{O}{3}]~4959,5007 emission is particularly strong and emission from
H$\alpha$ and [\ion{N}{2}]~6548,6583 is also detectable over a substantial area on
the sky.  The combination of emission line ratios, the close to zero
heliocentric radial velocity and the morphology of the structure are consistent
with an identification as a very nearby planetary nebula.  The proximity of the
hot, DO white dwarf PG~1034+001 further strengthens this interpretation.  The
object is: i) the largest planetary nebula on the sky, ii) certainly closer 
than any planetary nebula other than Sh 2--216, iii) the first to be 
unambiguously associated with a DO white dwarf. A parallax distance for 
PG~1034+001 would establish whether the structure is in fact the closest, and
one of the physically largest, planetary nebula known. 

\end{abstract}

\keywords{ISM: planetary nebulae: individual: (Hewett 1)---Stars: white dwarfs: individual 
(PG1034+001}

\section{Introduction}
\label{Intro}
The availability (Abazajian \etal 2003) of a significant fraction of the spectra and images from the
Sloan Digital Sky Survey (SDSS) (York \etal 2000) provides a unique resource
for the investigation of a wide variety of astrophysical phenomena.  The quality
and scale of the database are such that a number of serendipitous discoveries 
can be expected in the coming years. In this {\em Letter} we report the 
discovery, using SDSS spectra of unrelated objects, of the largest known 
planetary nebula (PN) on the sky.

\section{Data}

In the course of a search for SDSS spectra that show the signature of two objects,
at different redshifts, the distinctive presence of the [\ion{O}{3}]~4959,5007
doublet, at essentially rest--frame wavelength, in several adjacent spectra was noted.  A
more targeted search employed a simple 41--pixel ($57\,$\AA) median filter to
generate a ``continuum'' which was then subtracted from the original spectrum to
produce a ``difference'' spectrum.  Emission lines were identified in individual
and composite difference--spectra using standard matched--filter techniques.
The search revealed the presence of [\ion{O}{3}]~4959,5007 in more than 100
spectra with the flux in the [\ion{O}{3}]~5007 line ranging from $1.6 \times
10^{-15} {\rm erg}\,{\rm s}^{-1} {\rm cm}^{-2}$ down to the limit of
detectability of $\simeq 8 \times 10^{-17} {\rm erg}\,{\rm s}^{-1} {\rm
cm}^{-2}$. Surface brightnesses, per square arcsecond, can be obtained from
the fluxes measured in the spectra by dividing by the fibre area 
($7.1\,{\rm arcsec}^2$).

The detections were confined to objects in a region several degrees across
centered at approximately $\alpha = 10^{h} 37^{m}$ $\delta = -00\arcdeg 18'$ or 
$159.3^{\circ}$,$-0.3^{\circ}$ (J2000).  H$\alpha$,
H$\beta$, [\ion{N}{2}]~6548,6583 and [\ion{S}{2}]~6718,6732 emission lines were
also present in spectra in the same region.  Figure 1 shows the
spatial distribution of the spectra with detectable [\ion{O}{3}]~4959,5007
($\bullet$), H$\alpha$ ($\circ$), and [\ion{N}{2}]~6583 ($\times$).  The hatched
area, extending to a radius of $1\arcdeg$ from $10^{h} 37^{m}$ $-00\arcdeg
18'$, indicates a region where composite spectra, derived using groups of 25
spectra without individual [\ion{O}{3}]~4959,5007 detections, show unambiguous
evidence of [\ion{O}{3}]~4959,5007 emission.  Positions of objects with SDSS
spectra for which no individual detections were obtained are also indicated
($.$).

Figure 2 shows the wavelength regions containing [\ion{O}{3}]~4959,5007 and
H$\alpha$, [\ion{N}{2}]~6548,6583 plus [\ion{S}{2}]~6718,6732 for a composite,
continuum--subtracted, spectrum made using all the galaxy and quasar spectra
within $0.5\arcdeg$ of $10^{h} 37^{m}$ $-00\arcdeg 18'$ (upper panels).  Another
composite, using objects from a more distant arc--shaped region to the
south--west is also shown (lower panels).  The arc--shaped region is defined by
radial distance $0.7\arcdeg-1.3\arcdeg$ from $10^{h} 37^{m}$ $-00\arcdeg 18'$
and angular extent $180\arcdeg \le {\rm PA} \le 315\arcdeg$ relative to the same
position.  The composite from the central region displays very strong
[\ion{O}{3}]~4959,5007 and, while H$\alpha$ and [\ion{N}{2}]~6548,6583 are
clearly visible, shows the relative weakness of the hydrogen lines.  The
composite spectrum from large radii to the south--west illustrates the fall--off
in the strength of the [\ion{O}{3}]~4959,5007 emission with radius and the
marked variation in the H$\alpha$/[\ion{N}{2}]~6548,6583 ratio with position.
The spectra of stars were not included in the generation of the composite
spectra.  To avoid contamination of the composite spectrum from unrelated
emission and absorption features, wavelength regions associated with strong
emission and absorption features in the rest--frames of the galaxy and quasar
spectra were excluded from the construction of the composite spectra.

Examination of the sky--subtracted sky spectra in SDSS plates 273 and 274, which
contain the region, show no evidence for absorption at the location of any of
the emission lines.  The lack of absorption confirms that the emission line
fluxes associated with the $\simeq 2\arcdeg$ region, centered on $10^{h} 37^{m}$
$-00\arcdeg 18'$, are not affected significantly by more extended emission on
scales of $\sim 5\arcdeg$.

The resolution of the SDSS spectra ($\simeq 3.1$\AA) precludes a
reliable determination of the radial velocity. The centroids of the 
[\ion{O}{3}]~4959,5007, H$\alpha$ and [\ion{N}{2}]~6583 emission lines in
a composite of the 38 spectra showing the strongest [\ion{O}{3}]~4959,5007
emission give a heliocentric radial velocity of $-5\pm5\,{\rm km \, s^{-1}}$.
However, the amplitude is comparable to the wavelength accuracy of the SDSS
spectra and variations of several tens of ${\rm km \, s^{-1}}$ are evident
from spectrum to spectrum. In summary, the heliocentric velocity of the
gas is consistent with a value of $0\pm20\,{\rm km \, s^{-1}}$.

The large angular extent and small radial velocity suggest a relatively local
origin for the ionized gas.  The Galactic coordinates, $l=248, b=+48$, further
suggest that the gas is within a few hundred parsecs if the object lies within
the Galactic disc. The weakness of the hydrogen lines and the lack of any
bright early--type stars in the vicinity of the nebula rule out identification
as an \ion{H}{2} region.

\section{Further Investigations}

Narrow--band imaging of the [\ion{O}{3}]~4959,5007 lines ($\lambda_c = 5008$\AA,
$\Delta\lambda = 100$\AA) and the H$\alpha$+[\ion{N}{2}]~6548,6583 lines
($\lambda_c = 6568$\AA, $\Delta\lambda = 95$\AA) was undertaken using the Wide
Field Camera (WFC) on the Isaac Newton Telescope (INT) on the nights of 2003 May
1 and 21--27.  Exposure times of $900-1200\,$s (H$\alpha$) and $ 3 \times
900\,$s [\ion{O}{3}] were used, with shorter exposures of 600s for companion
broad--band images in $g$ and $r$ passbands.  An area of roughly a square
degree was imaged in both [\ion{O}{3}] and H$\alpha$.
After processing through the INT WFC pipeline (Irwin \& Lewis 2001) there was a
clear detection of complex [\ion{O}{3}] and H$\alpha$ nebulosity extending over
the whole region, confirming the reality of the spectroscopic detections.
The resulting continuum--subtracted, stacked [\ion{O}{3}] and
H$\alpha$ images are shown in Figure 3.  The area visible in the images is
indicated by the dashed outline in Figure 1.  The centrally concentrated
distribution of [\ion{O}{3}]~4959,5007 emission is particularly striking.  
A distinctive feature present in the [\ion{O}{3}] image is the arc--like 
structure visible at center--right. 

%\clearpage

\begin{table*}[t]
\centering
\small
\caption{Emission Line Fluxes in the $<0.5\arcdeg$ Composite Spectrum}

\begin{tabular}{ccc}

\hline
\hline

Species  & Wavelength & Flux \\
         &({\AA}ngstroms) & ($10^{-17} {\rm erg}\,{\rm s}^{-1} {\rm cm}^{-2}$) \\

\hline

{[\ion{Ne}{3}]}   & 3870 & $6.5\pm1.0$ \\
H$\beta$ & 4861 & $1.0\pm0.8$ \\
{[\ion{O}{3}]}    & 4959 & $10.6\pm 1.5$ \\
{[\ion{O}{3}]}   & 5007 & $38.6\pm 3.0$ \\
{[\ion{N}{2}]}    & 6548 & $1.8\pm 0.5$ \\
H$\alpha$       & 6563 & $7.7\pm 1.2$ \\
{[\ion{N}{2}]}    & 6583 & $5.1\pm 1.0$ \\
{[\ion{S}{2}]}    & 6717 & $1.7\pm 0.7$ \\

\hline

\end{tabular}
\end{table*}

\clearpage

In the $<0.5\arcdeg$ composite (Figure 2a,b) the [\ion{O}{3}]
$\lambda$5007/$\lambda$4959 ratio of $3.6\pm0.4$ and the [\ion{N}{2}]
$\lambda$6583/$\lambda$6548 ratio of $2.8\pm0.3$ are both consistent with their
theoretical values, which depend only on atomic physics for the conditions
pertaining in PN, suggesting the spectrum provides useful diagnostic information
(although new spectra of the high surface brightness features are needed).
H$\beta$, which is clearly present in individual spectra with the strongest
emission line fluxes, is barely detectable in the $<0.5\arcdeg$ composite
spectrum.  The H$\alpha$/H$\beta$ ratio of $7.5\pm3.5$ indicates some reddening
(formally A$_V$ $=$ 0.9$^{+0.3}_{-0.6}$).  Relatively weak underlying absorption
in the Balmer lines would help explain the low H$\beta$ flux while making little
difference to the strength of H$\alpha$.  The nearly reddening independent
ratios of [\ion{N}{2}]~$\lambda$6583/H$\alpha$ $=$ $0.8\pm0.2$ and
[\ion{N}{2}]~$\lambda$6583/[\ion{S}{2}]~$\lambda$6716 $=$ $2.9\pm1$ are both
within the ranges observed in nitrogen-rich PN.  The [\ion{O}{3}]
$\lambda$5007/H$\beta$ ratio of $38\pm10$ is very high, although not
unprecedented in PNs (and is biased high by the selection process).  Further
support for the classification as a PN comes from the detection of
[\ion{Ne}{3}]~3869.  The absence of detectable emission from lines of
\ion{He}{1} and \ion{He}{2} is consistent with normal abundances and the S/N of
the composite spectrum.  In summary, the emission line properties of both
composite and individual spectra are consistent with the properties of a PN with
a relatively hot central star.

Indeed, the images in the atlas of ancient PNs (Tweedy \& Kwitter
1996) contain some strikingly similar structures.  The centrally concentrated
morphology evident in the distribution of [\ion{O}{3}]~4959,5007 emission is
common and large variations in emission line ratio, including strong
[\ion{N}{2}] and [\ion{S}{2}] towards the outer edges of old PNs,
due to interactions between the ejecta and the ambient interstellar medium, are
often seen.  The most unusual property of the structure reported here is the
angular size, which, at $> 2\arcdeg$ diameter exceeds that of Sh 2--216, also long
considered the closest PN, with an angular size of $1.6\arcdeg$
(Fesen, Blair, \& Gull 1981; Tweedy \& Napiwotzki 1992).

Since there was no plausible ionizing source in the SDSS photometric catalogue,
the APM sky survey catalogues were used for further investigation of potential
ionizing sources of radiation.  In a $1.5\arcdeg$ region centered on $10^h 37^m$
$-00\arcdeg 20\arcmin$ one bright candidate ionizing source stood out at $10^{h}
37^{m} 03.875^{s}$ $-00\arcdeg 08\arcmin 19.59\arcsec$ J2000 (Epoch 1982), with
UKST plate magnitudes and colors of ${\rm R} = 13.06$ and ${\rm B_J - R} =
-0.47$ (see Figure 4).  Using the UKST plate as a reference frame the refined
APM POSS1 (Epoch 1952) position of this source is $10^{h} 37^{m} 04.047^{s}$
$-00\arcdeg 08\arcmin 20.53\arcsec$ J2000, giving a high proper motion of
$-86\pm 5$, $+31\pm 5\,{\rm mas}\,{\rm yr}^{-1}$.  The color, magnitude and
proper motion suggested the object was probably a hot WD at around $\approx$100
pc and a search, utilizing Simbad, reveals the source is PG~1034+001, a bright,
$V=13.23$ (Landolt 1992), DO white dwarf.  A very crude kinematic age of $\sim
130,000\,$yr, can be derived from the radius of the PN, $\simeq 1\arcdeg$ at a
distance of $\simeq$150pc, and a typical expansion velocity of PN, $\simeq
20\,{\rm km \, s^{-1}}$.  The WD is currently close to the center of the region
over which detectable emission is present in the SDSS spectra.  The WD is also
in close proximity, $\approx$15\arcmin, to the region showing the strongest
[\ion{O}{3}]~4959,5007 emission, however, the proper motion vector for the WD
indicates a position for the WD to the east of the present location at the
time the PN was formed.  If the arc--like feature visible in Figure 3 represents a shock, or a
boundary, associated with material ejected at the time the PN was formed then a
lower limit to the kinematic age for the PN can be estimated by extrapolating
the motion of the white dwarf back to the origin of the radius of curvature of
the arc.  The resulting limit on the age, $>50,000\pm10,000\,$yr is plausible
but an understanding of the nature of the arc--like feature is necessary to
validate the argument.  An age of $\approx$100,000yr is consistent with all the
evidence but measurement of the expansion velocity and a more complete mapping
of the morphology of the emission will provide a much improved estimate.

\section{Discussion}

Given that all observations are consistent with a PN nature for the newly
discovered nebulosity, we designate it Hewett~1.  The proximity of the DO white
dwarf PG~1034+001 (Wesemael, Green, \& Liebert 1985) makes it probable that
PG~1034+001 is responsible for the ionizing radiation.  Werner, Dreizler, \&
Wolff (1995) derived a spectroscopic distance for PG~1034+001 of
$155^{+58}_{-43}\,$pc and confirmation of the association would make Hewett~1
one of the closest PNs known (Napiwotzki 2001), the $2\arcdeg$
diameter corresponding to a physical size of $3.5-7.0\,$pc for the likely
distance range of $100-200\,$pc.  Again, Sh 2--216 provides the current
benchmark, with a trigonometric distance of $130^{+9}_{-8}\,$pc (Harris \etal
1997) and a physical size of $\simeq 3.5\,$pc.  A parallax determination for
PG~1034+001 would establish whether Hewett~1 is even closer.

Notwithstanding targeted searches (Werner \etal 1997), very few PNs are known to
be associated with non--DA (hydrogen poor) white dwarfs.  Hewett~1 may be the
first PN to be discovered associated with a DO white dwarf, although PG~0108+101
(Reynolds \etal 1987) and PG~0109+111 (Werner \etal 1997) have candidate
nebulae, neither detection is regarded as secure.  Since DO white dwarfs are
thought to evolve from PG~1159 stars (see Dreizler \& Werner 1996, and
references therein), it will be of interest to compare the properties of
Hewett~1 both to the PNs associated with PG~1159 stars (Napiwotzki 1999) and
theoretical models (G{\' o}rny \& Tylenda 2000).  The derivation of a reliable
age for the PN associated with PG~1034+001 will be of particular interest for
constraining the time scales associated with the late--stages of evolution of
post-asymptotic giant branch stars and the origin of PG~1159 stars and
helium--rich WDs.

\acknowledgments

EDS is grateful for the hospitality of the IoA during his sabbatical visit and
we thank Alan McConnachie for assisting with the INT imaging observations.  
We are grateful to the anonymous referee who provided a detailed review of
the original manuscript. This
paper includes observations made with the Isaac Newton Telescope, operated on
the island of La Palma by the Isaac Newton Group in the Spanish Observatorio del
Roque de los Muchachos of the Instituto de Astrofisica de Canarias.  Funding for
the Sloan Digital Sky Survey (SDSS) has been provided by the Alfred P.  Sloan
Foundation, the Participating Institutions, the National Aeronautics and Space
Administration, the National Science Foundation, the U.S.  Department of Energy,
the Japanese Monbukagakusho, and the Max Planck Society.  The SDSS Web site is
http://www.sdss.org/.

%The SDSS is managed by the Astrophysical Research Consortium (ARC) for the Participating Institutions.
%The Participating Institutions are The University of Chicago, Fermilab, the Institute for Advanced Study,
%the Japan Participation Group, The Johns Hopkins University, Los Alamos National Laboratory, the
%Max-Planck-Institute for Astronomy (MPIA), the Max-Planck-Institute for Astrophysics (MPA), New
%Mexico State University, University of Pittsburgh, Princeton University, the United States Naval
%Observatory, and the University of Washington.

\clearpage
%------------------------- figure 1 begins ------------------------------
{\vskip 0.5truecm 
\epsfxsize=8.5truecm 
\epsfbox{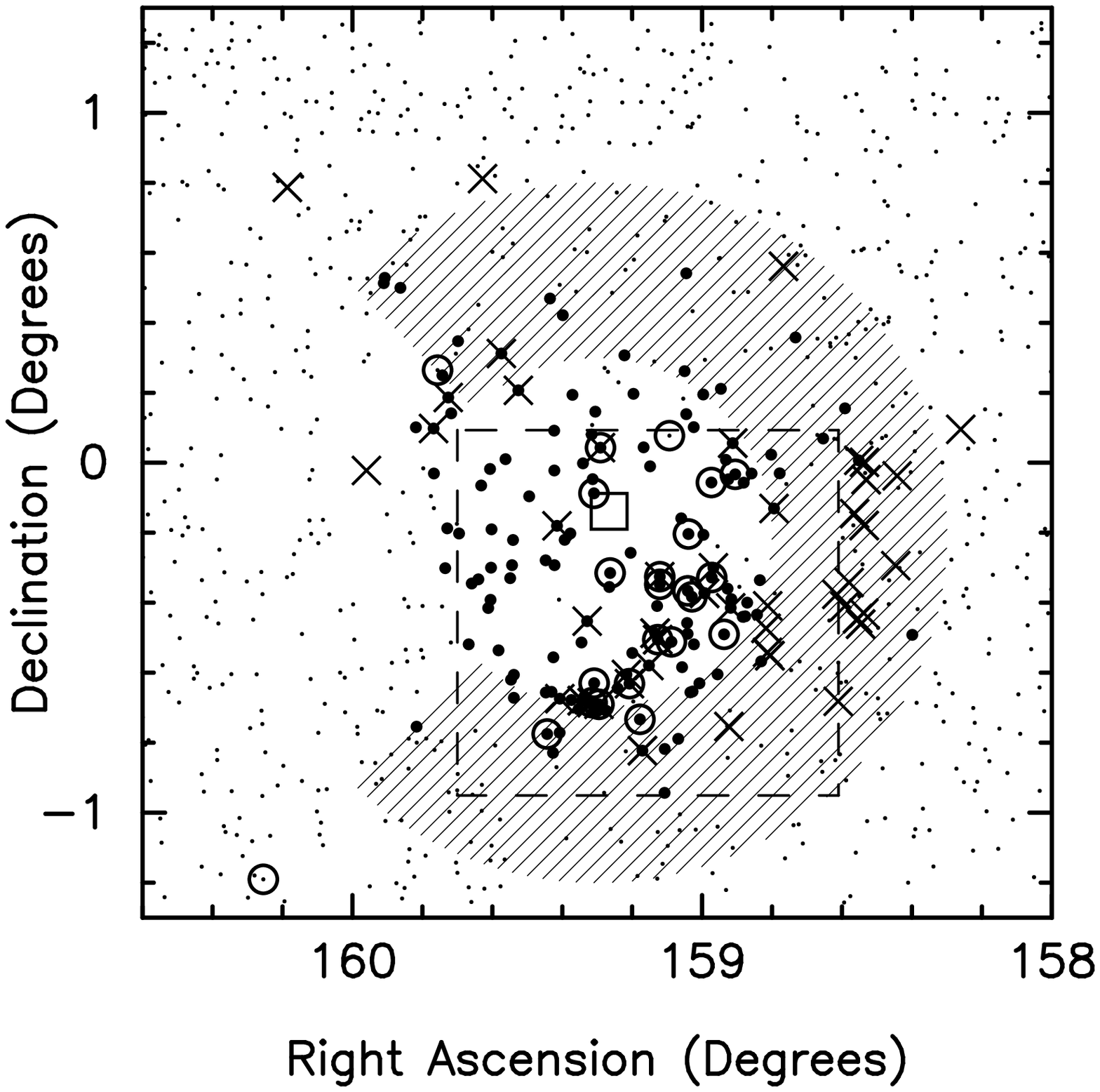} 
\figcaption[f1.eps]{\label{fig1}
{Spatial distribution of spectra with detectable [\ion{O}{3}]~4959,5007
($\bullet$), H$\alpha$ ($\circ$), and [\ion{N}{2}]~6583 ($\times$).  The 
hatched area indicates a region where composite spectra also show unambiguous
evidence of [\ion{O}{3}]~4959,5007 emission.  Positions of objects with SDSS
spectra for which no individual detections were obtained are also indicated
($\cdot$). The dashed outline shows the area included in the narrow--band images
of Figure 3. The location of the white dwarf PG~1034+001 is marked by a $\Box$.
}}
\vskip 0.5truecm}
%------------------------- figure 1 ends ------------------------------

\clearpage

%------------------------- figure 2 begins ------------------------------
\begin{figure*}[t]
\begin{center}
\leavevmode
\hbox{%
\epsfxsize=12.7cm
\epsffile{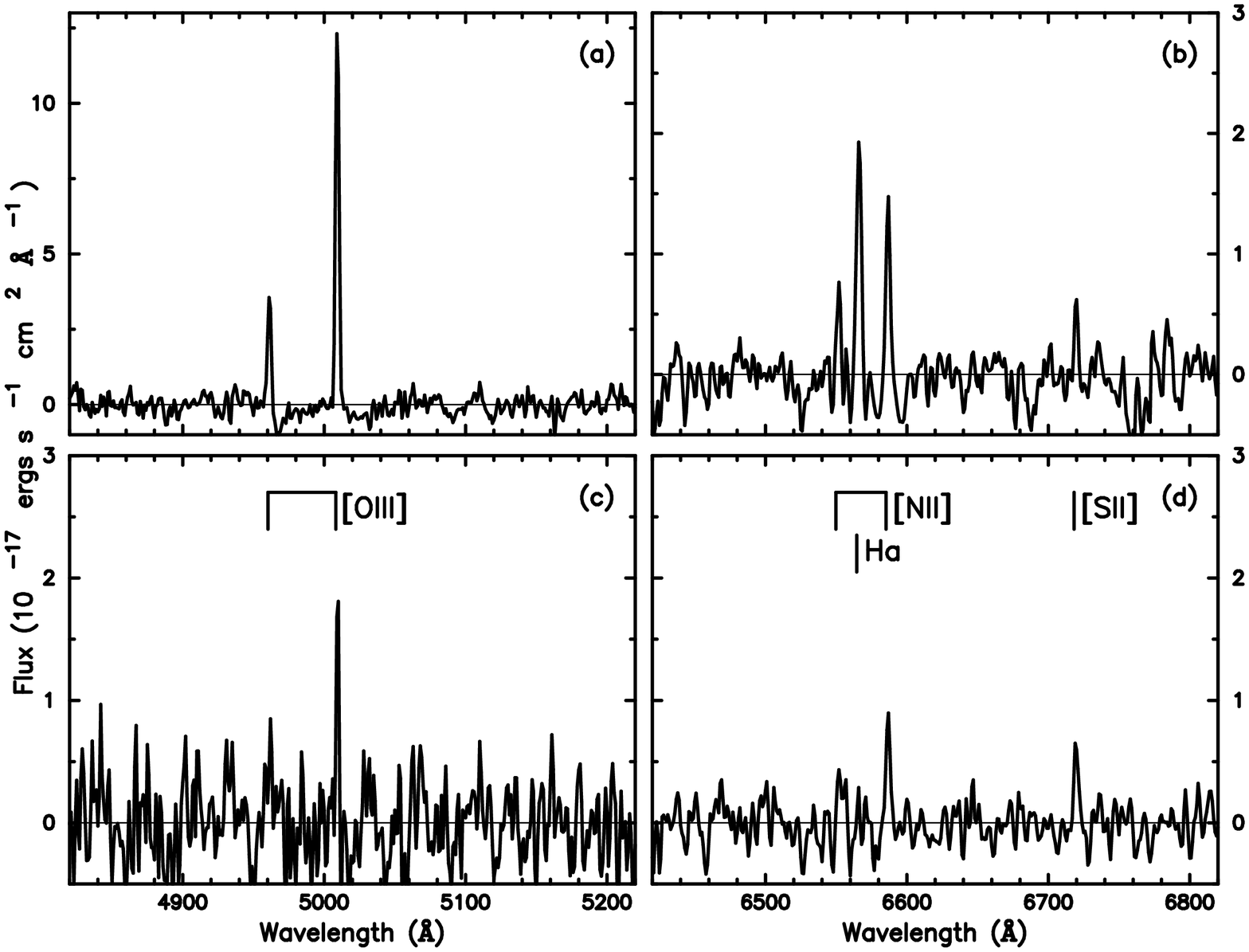}}
\caption[f2.eps]{\label{fig2}
{
Wavelength regions containing [\ion{O}{3}]~4959,5007 and
H$\alpha$,[\ion{N}{2}]~6548,6583 plus [\ion{S}{2}]~6718,6732 for a composite,
continuum--subtracted, spectrum made using all the galaxy and quasar spectra
within $0.5\arcdeg$ of $10^{h} 37^{m}$ $-00\arcdeg 18'$ (upper panels).
Another composite, using objects with position angles $180\arcdeg \le {\rm PA}
\le 315\arcdeg$ lying between $0.7\arcdeg-1.3\arcdeg$ from $10^{h} 37^{m}$
$-00\arcdeg 18'$ is also shown (lower panels).  The composite from the central
region displays very strong [\ion{O}{3}]~4959,5007 and, while H$\alpha$ and
[\ion{N}{2}]~6548,6583 are clearly visible, shows the relative weakness of the
hydrogen lines.  The composite spectrum from large
radii to the south--west illustrates the fall--off in the strength of the
[\ion{O}{3}]~4959,5007 emission with radius and the marked variation in the
H$\alpha$/[\ion{N}{2}]~6548,6583 ratio with position.
}}
\end{center}
\end{figure*}

%------------------------- figure 2 ends ------------------------------

\clearpage

%------------------------- figure 3 begins ------------------------------
\begin{figure*}[t] 
\begin{center} 
\leavevmode \hbox{% 
\epsfxsize=12.4cm
\epsffile{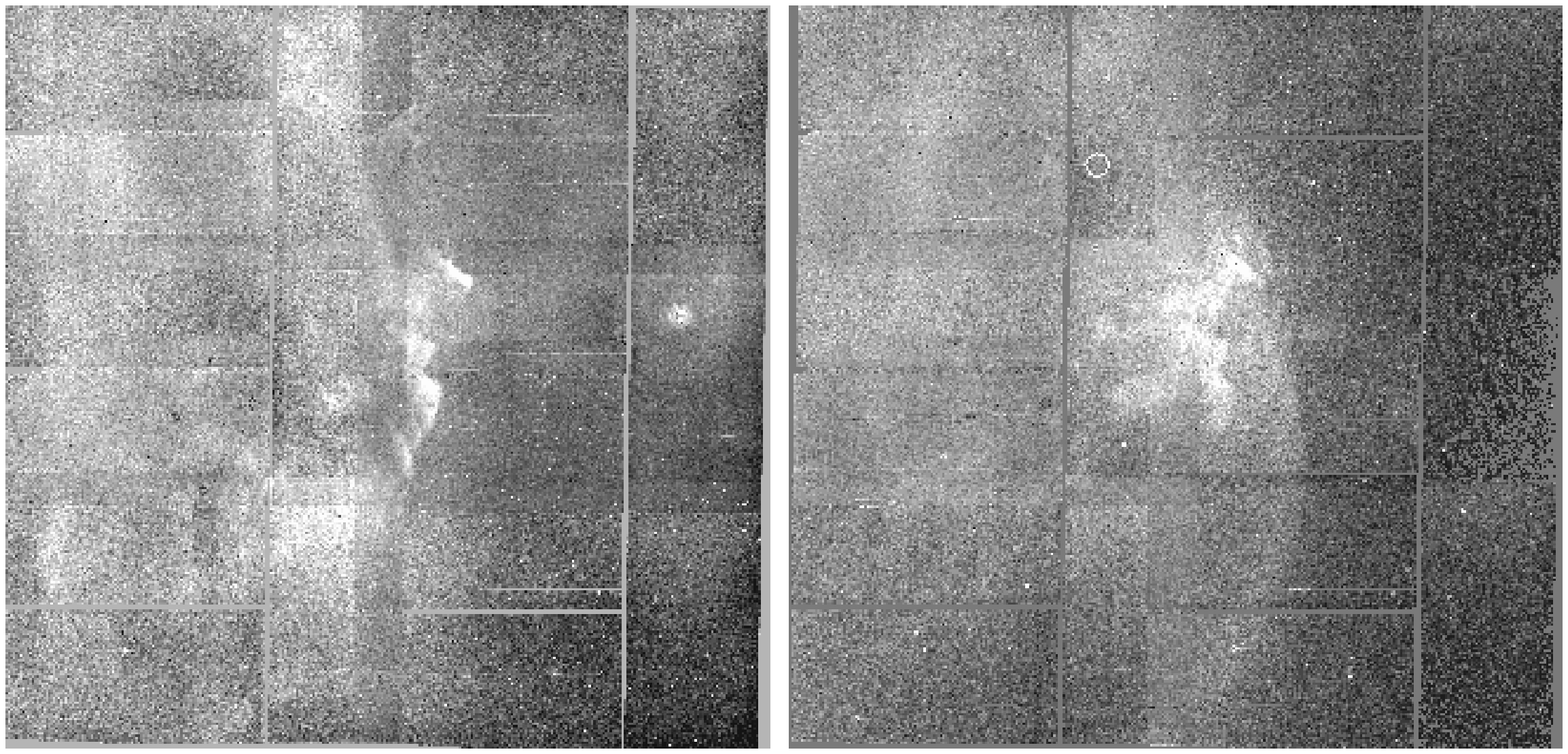}} 
\caption[f3.eps]{\label{fig3} 
{The left hand panel shows a mosaic of 6 continuum--subtracted pointings in
H$\alpha$+[\ion{N}{2}] while the right panel shows the equivalent for
[\ion{O}{3}].  The images are approximately $0.8^{\circ}$ on a side with North
to the top and East to the left.  The position of the white dwarf PG~1034+001 is
indicated by a circle in the [\ion{O}{3}] image. Emission with
complex structure is evident in the central regions of the images in both
passbands.  A well--defined arc, or boundary, is visible at center--right in the
[\ion{O}{3}] image.}}
\end{center}
\end{figure*}

%------------------------- figure 3 ends ------------------------------

\clearpage

%------------------------- figure 4 begins ------------------------------
{\vskip 0.5truecm 
\epsfxsize=7.5truecm 
\epsfbox{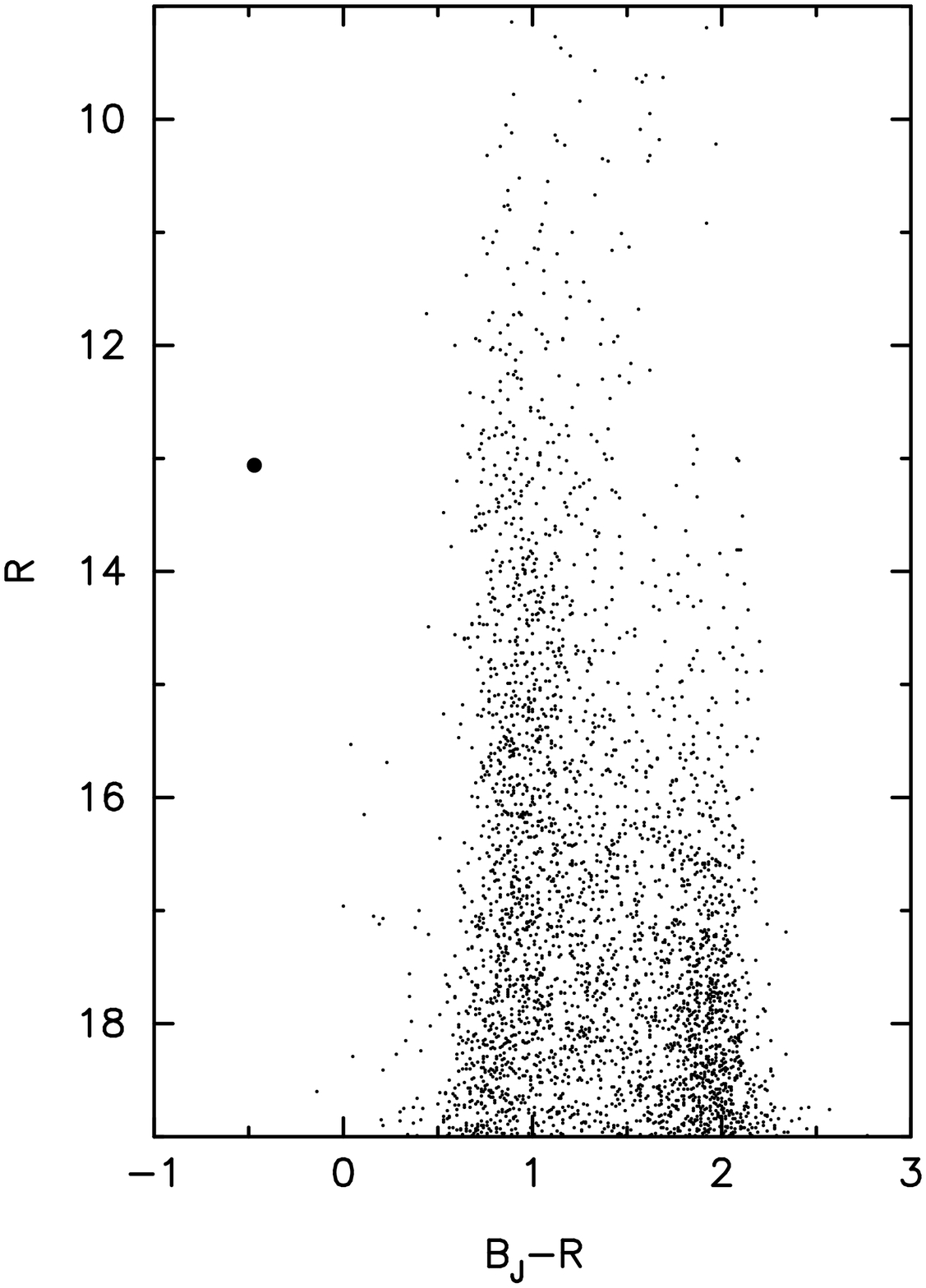} 
\figcaption[f4.eps]{\label{fig4}
{$B_J-R$ {\it versus} $R$ color--magnitude diagram for objects within
$0.75\arcdeg$ of $10^{h} 37^{m}$ $-00\arcdeg 18'$ using the APM sky catalogue
magnitudes of UK Schmidt Telescope $B_J$ and $R$ plates. All objects 
classified as stellar are plotted ($.$) with the location of the white--dwarf
PG 1034+001 highlighted ($\bullet$).}}
\vskip 0.5truecm}
%------------------------- figure 4 ends ------------------------------

\end{document}